\documentclass[aps,pre,reprint,amsmath,amssymb,superscriptaddress,showpacs,floatfix]{revtex4-2}

\usepackage{graphicx}
\usepackage{dcolumn}
\usepackage{bm}
\usepackage[colorlinks, allcolors=blue]{hyperref}
\usepackage[usenames, dvipsnames]{color}
\usepackage{float}

\usepackage{lipsum}
\usepackage{titlesec}
\titlespacing\section{0pt}{12pt plus 4pt minus 4pt}{1pt plus 20pt minus 2pt}
\usepackage{xcolor}
\usepackage{tabularx}
\usepackage{amsmath}
\usepackage{comment}
\usepackage{afterpage}
\usepackage{placeins}
\usepackage{booktabs}
\usepackage{multirow}
\usepackage{array}
\usepackage{setspace}
\graphicspath{{Figs/}}
\usepackage{siunitx}
\usepackage{hhline}
\usepackage{xfrac}
\usepackage{float,graphicx}
\usepackage{mathtools}
\usepackage{listings}
\usepackage{amssymb}
\usepackage[normalem]{ulem}
\usepackage{titlesec}
\usepackage{amsfonts}
\usepackage[version=4]{mhchem}

\usepackage{epstopdf}
\catcode`@11
\def\seceqaa{\@addtoreset{equation}{section}
principles\def\theequation{A\arabic{equation}}}
\def\seceqbb{\@addtoreset{equation}{section}
\def\theequation{B\arabic{equation}}}
\def\seceqcc{\@addtoreset{equation}{section}
\def\theequation{C\arabic{equation}}}
\def\seceqdd{\@addtoreset{equation}{section}
\def\theequation{D\arabic{equation}}}
\def\seceqee{\@addtoreset{equation}{section}
\def\theequation{E\arabic{equation}}}
\def\seceqff{\@addtoreset{equation}{section}
\def\theequation{F\arabic{equation}}}
\def\seceqgg{\@addtoreset{equation}{section}
\def\theequation{G\arabic{equation}}}
\def\seceqhh{\@addtoreset{equation}{section}
\def\theequation{H\arabic{equation}}}
\catcode`@11

\DeclareUnicodeCharacter{2009}{\,}
\newcommand{\SPINX}{\affiliation{Spin-X Institute, School of Physics and Optoelectronics, State Key Laboratory of Luminescent Materials and Devices, Guangdong-Hong Kong-Macao Joint Laboratory of Optoelectronic and Magnetic Functional Materials, South China University of Technology, Guangzhou 511442, China}}

\newcommand{\SINP}{\affiliation{Condensed Matter Physics Division, Saha Institute of Nuclear Physics, A CI of Homi Bhabha National Institute 1/AF, Bidhannagar, Kolkata, India}} 

\newcommand{\stu}{\affiliation{College of Integrated Circuits and Optoelectronic Chips, Shenzhen Technology University, Shenzhen 518118, Guangdong, China}}

\newcommand{\sustech}{\affiliation{Department of Physics, Southern University of Science and Technology, Shenzhen 518055, China}}

\newcommand{\IACS}{\affiliation{School of Physical Sciences, Indian Association for the Cultivation of Science, Jadavpur, Kolkata 700032, India}}

\newcommand{\PKUPHYS}{\affiliation{State Key Laboratory for Mesoscopic Physics, School of Physics, Peking University, Beijing 100871, China}}

\newcommand{\PKUMATER}{\affiliation{School of Materials Science and Engineering, Beijing Key Laboratory for Magnetoelectric Materials and Devices, Peking University, Beijing 100871, China}}

\newcommand{\KLNSTA}{\affiliation{Key Laboratory of Neutron Scattering Technology and Application, China National Nuclear Corporation, Beijing 100822, China}}

\newcommand{\SYSU}{\affiliation{School of Materials, Shenzhen Campus of Sun Yat-Sen University, Shenzhen 518107, China}}
\newcommand{\CIAE}{\affiliation{China Institute of Atomic Energy, Beijing 102413, China}}

\begin{document}

\title{
Magnetic field-induced non-trivial Lifshitz transition in TaCo$_2$Te$_2$}
\author{Suman Kalyan Pradhan}
\altaffiliation{These authors contributed equally to this work}
\email{suman1kalyan@scut.edu.cn}
\SPINX
\SINP

\author{Xiaoming Ma}
\altaffiliation{These authors contributed equally to this work}
\stu
\sustech 
\author{Jicheng Wang}
\altaffiliation{These authors contributed equally to this work}
\SPINX
\author{Weiqi Liu}
\SPINX

\author{Yue Dai}
\sustech

\author{Wenxing Chen}
\SPINX

\author{Xiaobai Ma}
\CIAE
\KLNSTA

\author{Wenyun Yang}
\PKUPHYS

\author{Yu Wu}
\PKUPHYS

\author{Zhaochu Luo}
\PKUPHYS

\author{Raktim Datta}
\IACS

\author{Arnab Bera}
\IACS

\author{Samik DuttaGupta}
\SINP

\author{Jinbo Yang}
\PKUPHYS

\author{Yanglong Hou}
\SYSU
\PKUMATER

\author{Chang Liu}
\email{liuc@sustech.edu.cn}
\sustech

\author{Rui Wu}
\email{ruiwu001@scut.edu.cn}
\SPINX

\date{\today}

\begin{abstract}
Magnetic-field-driven Lifshitz transitions are typically considered zero-temperature phenomena involving Fermi-surface reconstruction without symmetry breaking. Here, we report an unconventional Lifshitz transition in TaCo$_2$Te$_2$ that emerges exclusively within a narrow finite-temperature window under cooperative tuning by both temperature and magnetic field. Bulk-sensitive transport and thermoelectric measurements demonstrate continuous Fermi-surface renormalization at low temperatures, where the transition is sharply triggered by a critical magnetic field. Crucially, neutron diffraction reveals the absence of structural or magnetic phase transitions, while angle-resolved photoemission spectroscopy shows no spectral anomalies in electronic structure without magnetic field. These observations constrain the mechanism to a Zeeman-driven process invisible to equilibrium probes, establishing a paradigm where Fermi-surface topology is jointly controlled by temperature and magnetic field.

\end{abstract}
\maketitle
${Introduction-}$ 
Lifshitz transitions (LTs) \cite{Lifshitz} represent a distinctive class of quantum phase transitions in which the topology of the Fermi surface changes without any accompanying symmetry breaking or order-parameter formation \cite{Lifshitz,Varlamov}. Unlike conventional thermodynamic transitions, LTs arise from band-edge crossings relative to the chemical potential, leading to the creation or annihilation of Fermi pockets. They are therefore inherently electronic in origin and are most sharply defined at zero temperature, where they are typically accessed by tuning external parameters such as carrier density, magnetic field, pressure or strain \cite{Lifshitz, Varlamov, Liu2010, Pfau, Li,Aoki, Steppke2017, Wu, Zhang2017,Cai,Wu2023,Galeski}. As such topological reconstructions strongly reshape the low-energy density of states, Lifshitz transitions play a central role in a wide range of emergent quantum phenomena, including van Hove singularities, non-Fermi-liquid transport, electronic nematicity and unconventional superconductivity \cite{Steppke2017, Luo}. Moreover, the abrupt reconstruction of electronic topology near a LT leads to amplified transport responses, establishing a direct link between fundamental Fermi-surface physics and field-tunable electronic and thermoelectric functionalities.

Materials with small Fermi energies (Fermi level, $E_\text{F}$, lies very close to the band-touching points or lines that define the topology) and complex band dispersions are particularly susceptible to Lifshitz transitions, as even modest perturbations can induce pronounced changes in Fermi-surface topology. Topological semimetals provide a natural platform in this regard: their low carrier densities, symmetry-protected band degeneracies and proximity to band extrema make their electronic structure exceptionally sensitive to external tuning \cite{Qi, Hasan, Armitage}. Yet, despite extensive theoretical and experimental efforts, most realizations of Lifshitz transitions remain firmly rooted in a zero-temperature framework, with temperature largely treated as a source of thermal smearing rather than an active control parameter.

Within this context, the layered topological semimetal TaCo$_2$Te$_2$ has recently emerged as a promising system for exploring field-tunable electronic instabilities. TaCo$_2$Te$_2$ is air-stable and exfoliable, exhibits high carrier mobility and a large, non-saturating magnetoresistance, and hosts small electron and hole pockets arising from a nearly compensated electronic structure \cite{Singha, Wang, Jiao}. Magnetotransport measurements reveal unconventional carrier dynamics, including deviations from Kohler’s rule \cite{Pate}, while angle-resolved photoemission spectroscopy (ARPES) has identified multiple Fermi pockets, symmetry-protected band degeneracies and a high-multiplicity van Hove singularity \cite{Rong2023}. Quantum oscillation studies further indicate nontrivial Berry phases, underscoring the topological character of its electronic states \cite{Jiao}.

Despite this rich electronic landscape, a robust resistivity upturn emerges near 30 K under applied magnetic field \cite{Wang, Pate}, in the absence of any detectable structural distortion or long-range magnetic order. Its origin therefore remains ambiguous: whether it signals an incipient symmetry-breaking instability or a genuine reconstruction of the Fermi-surface topology. Importantly, a magnetic-field–induced resistivity upturn alone does not constitute definitive evidence for a Lifshitz transition, as similar anomalies may arise from orbital magnetoresistance or field-enhanced electronic correlations \cite{Lifshitz, Blanter1994, Pippard1989}. Resolving this distinction is intrinsically challenging, since field-driven LTs often involve subtle Fermi-surface pocket rearrangements that elude conventional probes.

We address this question through magnetotransport, thermoelectric, ARPES, and neutron diffraction measurements on exfoliated TaCo$_2$Te$_2$ flakes. The anomaly, confined to a temperature range of 20–30 K, is observed in Hall, planar Hall, and thermoelectric responses, while ARPES and neutron diffraction confirm the absence of structural or magnetic transitions. Thermoelectric signatures, sensitive to the energy derivative of the density of states, indicate a field-induced Lifshitz transition within this narrow temperature window. These results reveal a regime in which temperature and magnetic field jointly govern Fermi-surface topology in TaCo$_2$Te$_2$, extending the conventional zero-temperature framework for Lifshitz transitions and establishing this material as a model platform for disentangling topological electronic reconstructions from symmetry-breaking phenomena.
\begin{figure*}
\centering
\includegraphics[width=1.75\columnwidth]{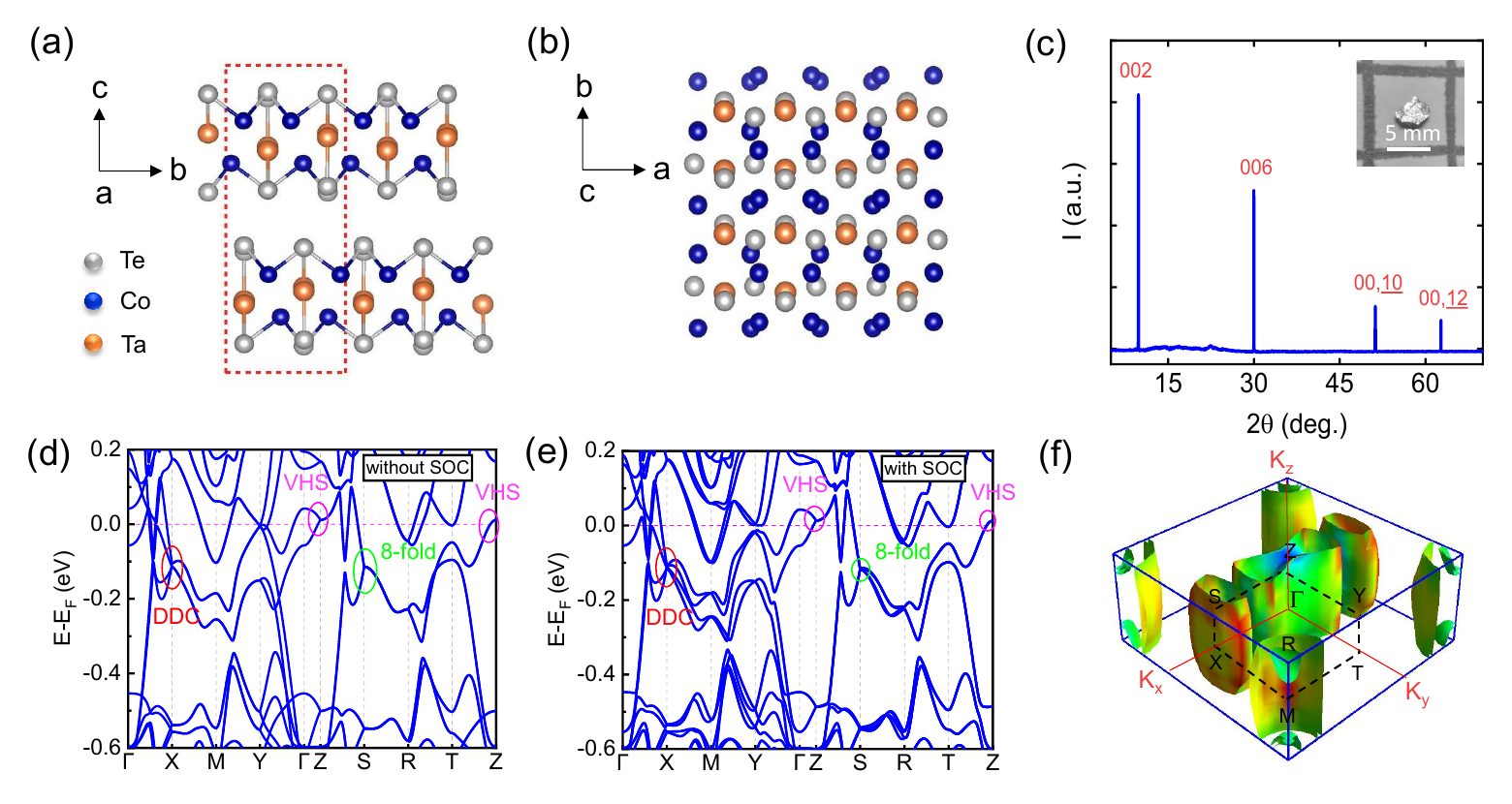}
\caption{
\textbf{Crystal structure and electronic properties of TaCo$_2$Te$_2$.} 
(a) The solved crystallographic unit cell of TaCo$_2$Te$_2$, composed of two monolayers stacked along the $c$ axis through weak van der Waals interactions; the unit cell is indicated by the red rectangle. 
(b) Crystal structure viewed along the $c$ axis. 
(c) Room-temperature X-ray diffraction (XRD) pattern of a TaCo$_2$Te$_2$ single crystal. Inset: photographic image of a representative grown crystal. 
(d,e) First-principles electronic band structures calculated along the high-symmetry path, shown (d) without and (e) with inclusion of spin–orbit coupling (SOC). 
(f) Three-dimensional Fermi surface topology obtained from SOC-inclusive calculations.}
\vspace{-0.45cm}
\label{DFT}
\end{figure*}

 ${Experimental}$ ${Results}$ \& ${Discussions-}$ For this work, high-quality millimeter-sized single crystals of TaCo$_2$Te$_2$ were synthesized using the chemical vapor transport (CVT) method, with full experimental details provided in the Supporting Information \cite{supple}. Single-crystal X-ray diffraction (SC-XRD) confirms that TaCo$_2$Te$_2$ crystallizes in the orthorhombic Pnma structure (space group No. 62) [Figs.~\ref{DFT}(a,b)], with lattice parameters $a = 6.6035$~\AA, $b = 6.5778$~\AA, and $c = 17.7762$~\AA. These experimentally determined lattice constants were used as the starting point for structural relaxation and subsequent first-principles calculations. Fig.~\ref{DFT}(c) shows the XRD patterns of the grown crystals, displaying only (00$l$) reflections, which indicates that the crystals are well-oriented along the $c$-axis \cite{Singha,Rong2023}. EDX results confirm the sample composition (Fig.~S1 \cite{supple}). The resulting band structure, shown in Figs.~\ref{DFT}(d,e), reveals that the inclusion of spin–orbit coupling (SOC) introduces only minimal changes, confirming that the low-energy electronic topology is primarily determined by the crystal symmetry and orbital character rather than SOC-induced band splitting. The calculated electronic structure reveals a double Dirac cone (DDC) at the X point and an eightfold-degenerate crossing at the S point, in agreement with recent reports \cite{Rong2023,Jiao}. A double Dirac cone consists of two coinciding Dirac cones, necessitating a fourfold-degenerate state at the Dirac point. In contrast to earlier studies \cite{Rong2023}, the van Hove singularity (VHS) at the Z point, in the density of states, (arise from stationary points in the quasiparticle energy dispersion, where $\nabla$E($\text{k}$) = 0 \cite{Van}), 
 is located above the Fermi level in our calculations, reflecting the theoretically determined chemical potential adopted here. To further elucidate the electronic topology of TaCo$_2$Te$_2$, we compute the full three-dimensional Fermi surface, shown in Fig.~\ref{DFT}(f).
 
\begin{figure*}
\centering
\includegraphics[width=1.7\columnwidth]{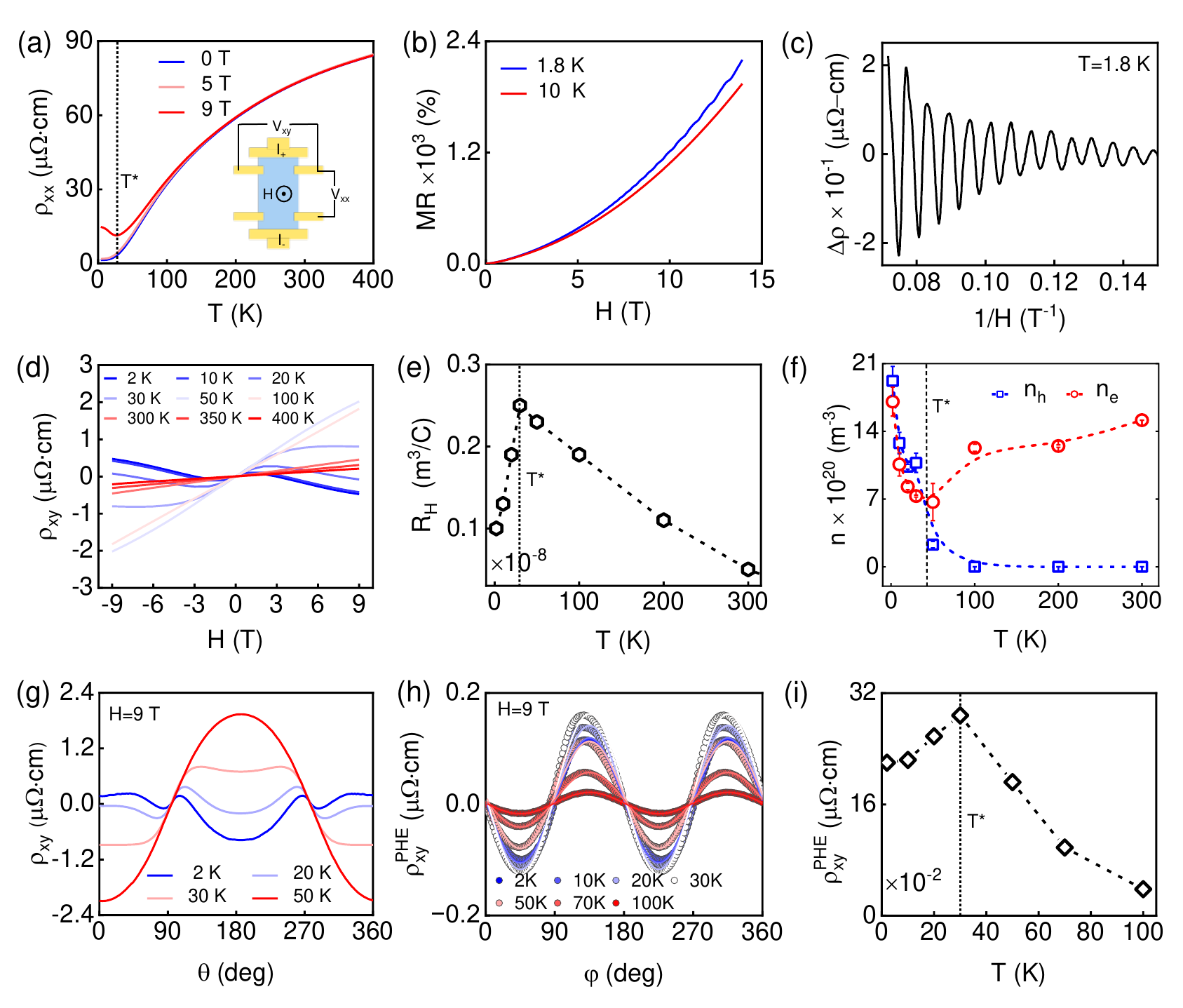}
\caption{\textbf{A distinct anomaly around 30 K in the transport measurements - points to a potential reconstruction of the Fermi surface :} 
Magneto-resistance - (a) Temperature dependence of resistivity measured under $H$ = 0, 5, and 9 T magnetic fields. Lower inset: schematic of a Hall bar device.
(b) The field-dependent magnetoresistance with the magnetic field perpendicular to the ${c}$ axis up to 14 T, at 1.8 and 10 K. (c) Details of the quantum
oscillations at 1.8 K, after background subtraction, demonstrating the high quality of the crystal.
Conventional Hall effect - (d) Field dependence of $\rho_{xy}(H)$ at various temperatures for the out-of-plane (OOP) configurations. (e) The temperature dependence of the Hall coefficient $R_\text{H}$ =  $d\rho_\text{xx}(H)/dH$ at the zero field limit.  
(f) Temperature dependence of the concentration for electron-type ($n_e$) and hole-type ($n_h$) carriers, obtained from Two-band model fitting. The lines serve as a guide to the eye.
(g) Angle ($\theta$)-dependent study of 
$\rho_\text{xy}$ at different temperatures under a magnetic field of 9 T. Planar
Hall effect - (h) Angular variation of $\rho^\text{PHE}_\text{xy}$, measured under 
different temperatures. The continuous line indicates
the fitted curve. 
(i) The temperature dependence of the PHE amplitude ($\rho^\text{PHE}_\text{xy}$), shown by hollow cubes, exhibits a pronounced maximum at 30 K. For these transport measurements, we used a flake with a thickness of about 51 nm.}
\vspace{-0.45cm}
\label{Hall}
\end{figure*}

 For transport and thermoelectric measurements, thin TaCo$_2$Te$_2$ flakes were prepared by mechanical exfoliation of bulk single crystals using the Scotch-tape method. A representative flake with a thickness of 51~nm was used in this study (Fig.~S2 \cite{supple}). Figure \ref{Hall}(a) presents the temperature dependence of the longitudinal resistivity $\rho_\text{xx}(T)$, which displays a metallic profile with no signature of structural or magnetic transitions under zero-field cooling. The high residual resistivity ratio (RRR $\approx$ 53) attests to the exceptional crystalline quality. From 2 to 150 K, the temperature dependence of $\rho_\text{xx}$ is accurately captured by the Bloch–Grüneisen model (Fig.~S3 \cite{supple}) \cite{Cvijović}, consistent with dominant electron–phonon scattering in this regime. Upon applying a magnetic field of 9 T, however, a subtle but reproducible resistivity plateau emerges near $T^\star$ $\approx$ 27 K, in agreement with previous reports \cite{Wang, Pate}. Such field-induced anomalies are frequently observed in topological semimetals when magnetic fields shift the chemical potential toward band-touching points or Lifshitz saddle points, thereby modifying the curvature of the Fermi surface without breaking symmetry \cite{Shekhar, Tafti2016}. As established later in this study, the anomaly at $T^\star$ constitutes the first transport signature of a field-driven Fermi-surface reconstruction.

Figure \ref{Hall}(b) shows the field dependence of the magnetoresistance (MR) for the out-of-plane (OOP) configuration at 1.8 K. Two prominent features emerge:(i) a large, non-saturating positive MR of $\approx$ 2.3 $\times$ 10$^3$\%, and (ii) clear Shubnikov–de Haas (SdH) oscillations at high fields [Fig.~\ref{Hall}(c)]. Such a large MR typically originates from high carrier mobility and/or near electron–hole compensation \cite{Ziman2001}, both of which are consistent with the quantum-oscillation data and the known multiband semimetallic character of TaCo$_2$Te$_2$. The MR follows a power-law field dependence but deviates from the quadratic $H^2$ behavior expected for perfectly compensated systems, reminiscent of topological materials with linearly dispersing fermions whose Landau quantization deviates from conventional semiclassical expectations \cite{Liang2015}. A pronounced MR anisotropy is also observed, reaching a ratio of nearly 26 at 2 K [Fig.~S4(a,b) \cite{supple}]. Comparable anisotropies have been reported in NbSb$_2$ \cite{Wang2014} and topological nodal-line semimetals such as ZrSiS \cite{Singha2017}. In TaCo$_2$Te$_2$, the strong anisotropy reflects the quasi-two-dimensional character of its Fermi surface \cite{Schoop2016}, in contrast to isotropic three-dimensional metals where the MR ratio is expected to approach unity \cite{Campbell1982Chapter9T}. This anisotropic response provides an additional indication that the applied magnetic field reshapes the Fermi-surface geometry—consistent with a Lifshitz-like instability.

\begin{figure*}
\centering
\includegraphics[width=1.7\columnwidth]{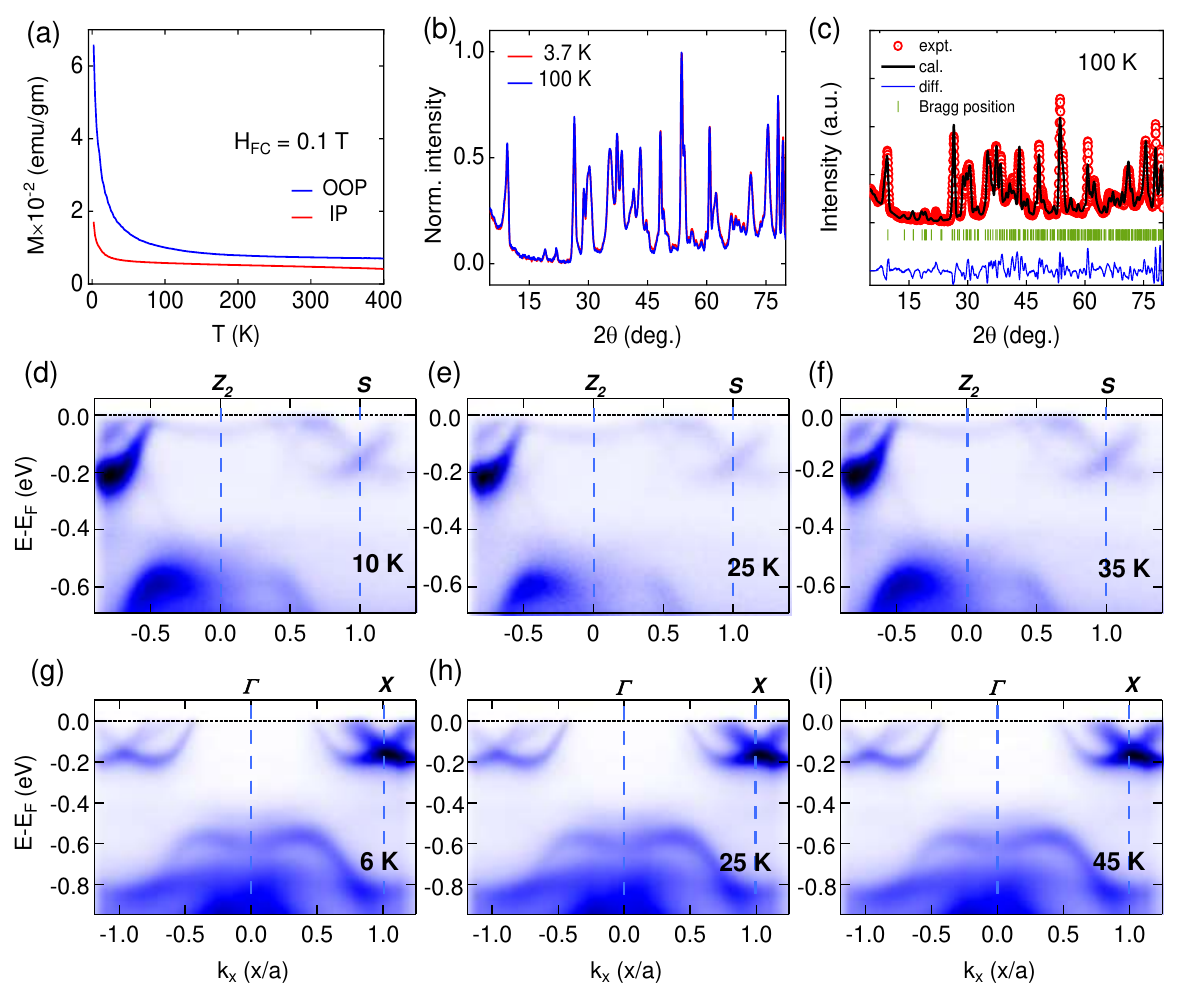}
\caption{
\textbf{Symmetry-conservation :} (a) Temperature dependence of the field-cooled (FC) magnetization under an external magnetic field of 0.1 T. 
(b) Temperature-dependent neutron powder diffraction (NPD) patterns collected at 3.7~K and 100~K.
(c) Rietveld refinement of the NPD data at 100~K, confirming the absence of structural or magnetic symmetry breaking.
(d–f) Temperature-dependent ARPES spectra along the Z$_2$–S direction measured at (d) 10~K, (e) 25~K, and (f) 35~K.
(g–i) ARPES spectra along the $\Gamma$–X direction at (g) 6~K, (h) 25~K, and (i) 45~K, showing no detectable band shifts across the measured temperature range.
}
\vspace{-0.45cm}
\label{ARPES}
\end{figure*}

Before discussing the Hall-effect signatures in detail, we first examine the magnetic response of the system. Temperature-dependent magnetization measurements [Fig.~\ref{ARPES}(a)] reveal no discernible anomalies across the entire investigated range, indicating the absence of any magnetic phase transition between 2 and 400 K. To further substantiate this conclusion, we performed neutron powder diffraction (NPD) measurements across the anomaly near 30 K, both below and above the putative transition. As shown in Fig.~\ref{ARPES}(b,c), the diffraction patterns exhibit no evidence of structural distortion or changes in magnetic symmetry. Together, these results establish that TaCo$_2$Te$_2$ preserves both crystallographic and magnetic symmetries across the anomaly. Since magnetic ordering would break time-reversal ($\mathcal{T}$) symmetry and reduce crystal symmetries, their absence confirms that the observed transport anomaly originate from a purely electronic reconstruction of the Fermi surface.

Building on the carrier dynamics inferred from the MR response, we performed detailed Hall effect measurements [Fig.~\ref{Hall}(d)], exploiting the high sensitivity of $\rho_{xy}(H)$ to Fermi-surface topology. 
At elevated temperatures (400--50 K), $\rho_{xy}(H)$ remains nearly linear, indicative of a dominant single-carrier contribution in this regime. Upon cooling below 50 K, however, $\rho_{xy}(H)$ develops pronounced low-field nonlinearity, signaling the progressive involvement of multiple carrier pockets and enhanced multiband transport \cite{Akiba,Chi2017,Chen}. A more striking evolution emerges below $T \approx 30$ K. In this regime, the initially positive slope of $\rho_{xy}(H)$ at low fields gradually reverses at higher fields, revealing a crossover toward nearly compensated semimetal behavior. Such a field-driven inflection in $\rho_{xy}(H)$ is a hallmark of an underlying reorganization of electron and hole pockets, commonly associated with changes in Fermi-surface curvature rather than symmetry breaking \cite{Chen}. This interpretation is further supported by the temperature dependence of the Hall coefficient $R_\text{H}(T)$ in the zero-field limit, which exhibits a clear hump at $\sim 30$ K [Fig.~\ref{Hall}(e)]. The hump marks the temperature where the effective carrier density reaches its minimum---a characteristic signature expected when a Lifshitz transition modifies the available phase space for charge carriers. The nonlinear Hall effect and the anomalous behavior of $R_\mathrm{H}$ together constitute compelling transport evidence for a magnetic-field–driven reconstruction of the Fermi surface in TaCo$_2$Te$_2$.\\
To further substantiate the transport signatures discussed above, we employed a semiclassical two-band model \cite{Chen} to analyze the field-dependent Hall response (Fig.~S5 \cite{supple}). The extracted carrier parameters reveal that the electron density consistently exceeds the hole density across the measured temperature range, in good agreement with earlier reports on TaCo$_2$Te$_2$ \cite{Singha,Wang}. As illustrated in Fig.~\ref{Hall}(f), both electron and hole density  ($n_e$ and $n_h$) decrease monotonically with increasing temperature, yet exhibit pronounced minima in the vicinity of 30 K. A similar nonmonotonic evolution is observed in the temperature dependence of the carrier mobilities ($\mu_e$ and $\mu_h$). Such simultaneous anomalies in both carrier concentration and mobility strongly indicate a reorganization of the low-energy electronic states. In particular, the sharp suppression and subsequent recovery of $n_e$ suggest the emergence of an additional electron pocket or a reshaping of the existing Fermi-surface contours near $T^\star$. This behavior is fully consistent with the onset of a Lifshitz transition, wherein subtle topological changes in the Fermi surface occur without breaking any underlying symmetry.

Furthermore, the angular dependences of the  Hall resistivity, $\rho_{xy}(\theta)$, undergo a distinct modification above 30 K [see Fig.~\ref{Hall}(g)]. Above 30 K, the way the Hall resistivity changes as the magnetic field is rotated becomes noticeably different [Fig.~\ref{Hall}(g)]. This suggests that the Fermi surface—the collection of electronic states at the Fermi level—undergoes a temperature-assisted change in its shape or dimensional character when a magnetic field is applied. In this situation, even small shifts of the energy bands caused by the field (Zeeman effect) can upset the balance between electrons and holes, causing different parts of the Fermi surface to contribute unevenly to the Hall conductivity $\sigma_\text{xy}$ and resulting in a significant reorganization of the Hall response.\\
\begin{figure}
\centering
\includegraphics[width=1\columnwidth]{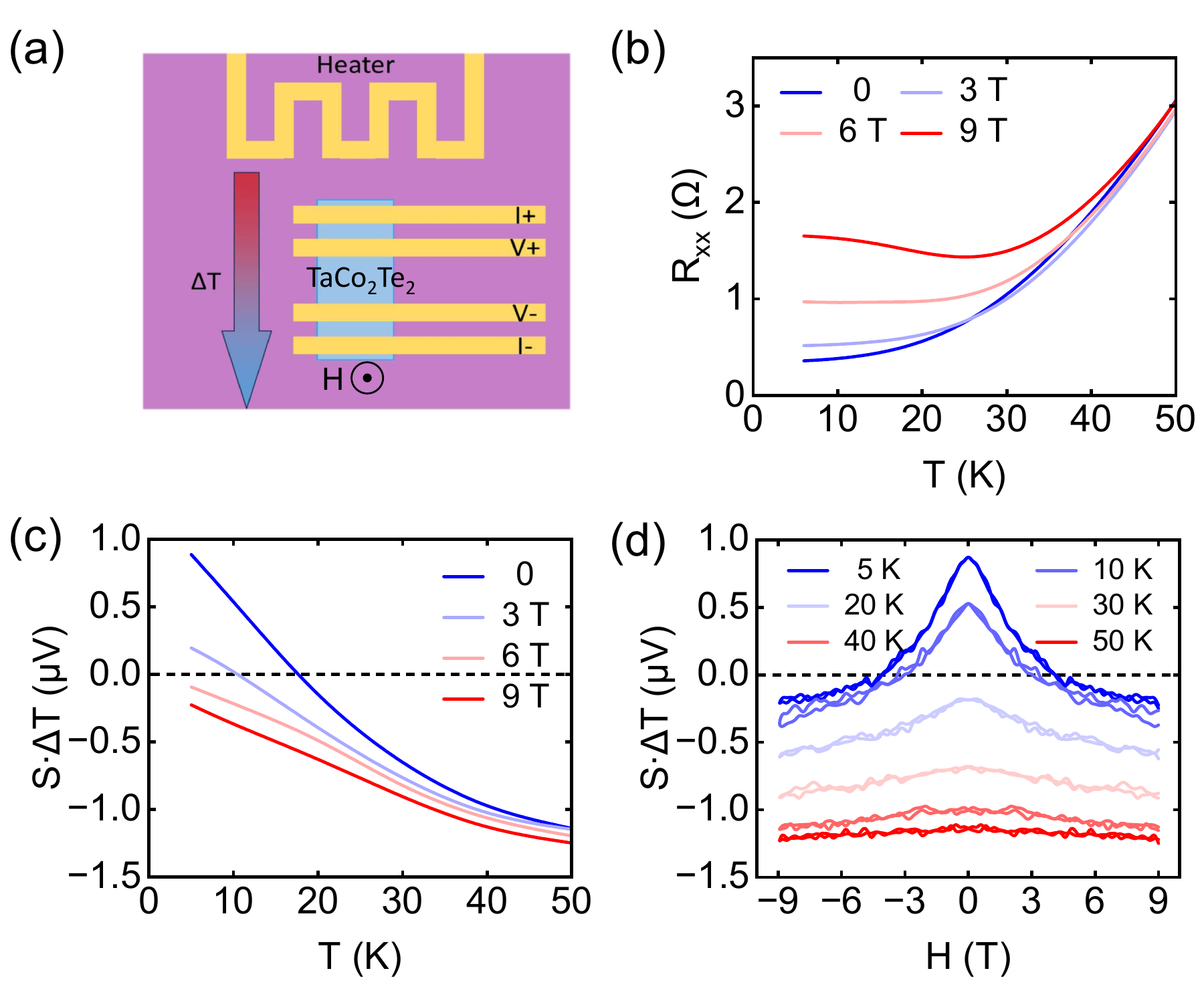}
\caption{\textbf{Thermelectric study :}
(a) Schematic diagram of the device for thermoelectric measurement. The sample is oriented with the magnetic field applied along the $c$ axis. The thickness of the flake is approximately 47~nm. (b) Temperature dependence of the resistivity under applied magnetic fields of $H = 0$, 3, 6, and 9~T.  (c) Temperature dependence of the Seebeck coefficient under different magnetic fields. (d) Seebeck coefficient $S$ as a function of magnetic field at selected temperatures.}
\vspace{-0.45cm}
\label{Thermoelectric}
\end{figure}
Beyond the conventional Hall effect, the planar Hall effect (PHE) provides an exceptionally sensitive probe of subtle electronic phase transitions in topological semimetals \cite{Liu2020}. In TaCo$_2$Te$2$, we observe a pronounced and unusually robust PHE signal that persists up to elevated temperatures [Fig.~\ref{Hall}(h)], highlighting its intrinsic electronic origin. The temperature dependence of the PHE amplitude displays a distinct maximum near 30 K [Fig.~\ref{Hall}(i)], coinciding precisely with anomalies in both the longitudinal resistivity $\rho_{xx}(T)$ and the Hall coefficient. The concurrent anomalies in the magnitude and angular evolution of the PHE are characteristic of a reconstruction of the Fermi-surface topology rather than a symmetry-breaking phase transition, providing compelling transport evidence for a Lifshitz transition in TaCo$_2$Te$_2$.

The anomalies observed in the magneto-transport properties of TaCo$_2$Te$_2$, including the nonlinear Hall response and deviations in $\rho_{xx}(T)$ and the Hall coefficient near 30 K, point toward an underlying electronic reconstruction. However, magneto-transport measurements alone cannot unambiguously distinguish between changes arising from carrier mobility, scattering mechanisms, or a genuine modification of the Fermi-surface topology \cite{Wang19}. This motivates the inclusion of thermoelectric measurements, which are intrinsically sensitive to the energy dependence of the electronic structure near the Fermi level ($S = \frac{8\pi^2 k_\mathrm{B}^2}{3 q h^2}\, m^* T \left(\frac{\pi}{3n}\right)^{2/3}$\cite{ioffe1960physics}, here, $k_\mathrm{B}$ denotes the Boltzmann constant, $q$ the carrier charge, $n$ the carrier concentration, $h$ the Planck constant, and $m^*$ the density-of-states effective mass). At fixed temperature, the Seebeck coefficient follows $S \propto (\pi/3n)^{2/3}$. Because $n$ directly controls the Fermi-level position, thermopower measurements serve as a powerful and bulk-sensitive probe of field-induced Lifshitz transitions.

Motivated by this, we measured the Seebeck coefficient $S$ as a function of temperature and magnetic field. The thermoelectric measurement setup is depicted in Fig.~\ref{Thermoelectric}(a), where a miniature heating wire was used to generate a longitudinal temperature gradient. The thermoelectromotive force ($\Delta V$) was measured using the standard four-point probe method, yielding $S=\Delta V/\Delta T$. Fig.~\ref{Thermoelectric}(b) presents the temperature dependence of the device longitudinal resistivity; the field-induced anomalies observed are consistent with the aforementioned results, with $T^{*}\sim25$~K. The magnetic-field dependence of $S$ is shown in Fig.~\ref{Thermoelectric}(c). The data reveal that $S$ decreases with increasing magnetic field strength, and that its magnetic-field sensitivity progressively weakens as the temperature increases. Notably, at low temperatures, $S$ undergoes a sign reversal with increasing magnetic field, indicating a switch in the dominant carrier type. This behavior is consistent with the nonlinearity observed in the previously measured $\rho_{xy}(H)$ curves, where the slope reverses sign between the low-field and high-field regimes. Further insight is provided by the thermopower data shown in Fig.~\ref{Thermoelectric}(d). The longitudinal Seebeck coefficient $S_{xx}(T)$ exhibits a monotonic temperature dependence, accompanied by a magnetic-field-induced polarity reversal below $T^{*}$. Near a Lifshitz transition, the Seebeck coefficient is expected to vary abruptly at low temperatures~\cite{Abrikosov2017,Varlamov1989b}, because the proximity of the Fermi level to a band critical point produces rapid, and often divergent, changes in the energy derivative of the density of states—particularly on the side where the number of Fermi-surface pockets is larger. Such a pronounced enhancement of the thermoelectric response is therefore a hallmark of a topological Fermi-surface reconstruction. Collectively, the zero-field and field-tuned behavior of $S$~\cite{Ito,Boukahil} establishes a coherent thermoelectric signature of a magnetic-field-driven Lifshitz transition in TaCo$_2$Te$_2$ around 25~K.

Taken together, the magnetotransport and thermoelectric responses point to a unique field-driven Lifshitz transition confined to a narrow temperature window. Direct ARPES measurements are therefore essential for probing the Zeeman-induced band evolution and the associated Fermi-surface reconstruction, as ARPES provides a direct, momentum-resolved view of the electronic structure that complements transport measurements \cite{Zhang2017}. To determine whether the transport anomalies near 30 K originate from intrinsic band-structure modifications, we conducted systematic temperature-dependent ARPES measurements on TaCo$_2$Te$_2$.

As shown in Fig.~S6 \cite{supple}, constant-energy contour maps and temperature-dependent band dispersions were acquired using photon energies of 24 and 20~eV. Assuming an inner potential of 14~eV, these photon energies correspond to $k_z \approx 17\pi/c$ and $16\pi/c$, respectively. The measured band structure is in good agreement with previously reported results~\cite{Rong2023}. The van Hove singularity located near the Fermi level at the $Z_2$ point and the double Dirac cone feature at the $X$ point are clearly resolved in Fig.~\ref{ARPES}(d) and Fig.~\ref{ARPES}(g), respectively. To track the evolution of the van Hove singularity with temperature, ARPES measurements at a photon energy of 24~eV were performed in the second Brillouin zone along the $Z_2$–$S$ high-symmetry direction. As shown in Figs.~\ref{ARPES}(d-f), no discernible changes in the band structure are observed over the temperature range from 10 to 35~K. Complementary measurements at 20~eV probing the $\Gamma$–$X$ direction in the first Brillouin zone [Figs.~\ref{ARPES}(g-i)] similarly reveal temperature-independent band dispersions from 6 to 45~K. Therefore, ARPES measurements reveal no discernible band shifts or Fermi-surface reconstructions, indicating that the zero-field electronic structure remains unchanged across the anomalous temperature range.

A defining hallmark of a Lifshitz transition is the tuning of a van Hove singularity through the Fermi level, marking a change in Fermi-surface topology without symmetry breaking [Fig.~\ref{DFT}(d,e)]~\cite{Mori2019,Marques}. In TaCo$_2$Te$2$, ARPES resolves a saddle-point–derived van Hove singularity located in close proximity to the Fermi energy [Figs.~\ref{ARPES}(d)], placing the system near a topological instability of the electronic structure. This proximity implies that an applied magnetic field can readily shift a critical band extremum across $E_\text{F}$, thereby driving a Fermi-surface reconstruction. The associated enhancement of the density of states naturally accounts for the pronounced thermoelectric and anisotropic transport responses, establishing the Lifshitz transition as a fundamentally electronic phenomenon. Within this picture, the magnetic field acts primarily through Zeeman splitting, selectively shifting spin-polarized bands while leaving the zero-field band structure essentially unchanged within the experimental resolution of ARPES. The transition therefore manifests only under finite magnetic field, consistent with its narrow temperature window and its absence in zero-field spectroscopy.

In summary, by combining bulk-sensitive transport and thermoelectric measurements with momentum-resolved ARPES and microscopic magnetic probes, we establish a unified picture of the Lifshitz transition in TaCo$_2$Te$_2$. While temperature continuously renormalizes the band structure, the topological reconstruction of the Fermi surface is triggered sharply by an applied magnetic field, without accompanying symmetry breaking. This separation of thermal tuning from field-induced topology identifies a Zeeman-driven mechanism that operates within a narrow temperature window and remains invisible in zero-field spectroscopy. TaCo$_2$Te$_2$ thus emerges as a model system for accessing magnetic-field–controlled Lifshitz physics at finite temperature, accompanied by the appearance of VHS, demonstrating how complementary experimental probes are essential for resolving subtle electronic instabilities. More broadly, our results point to a viable route for engineering Fermi-surface topology in quantum materials using external fields.

${Acknowledgment:}$ This work is supported by the National Key R\&D Program of China (grant nos. 2022YFA1203902, 2022YFA1403700), the National key research and development program of China (grant nos. 2022YFA1203904), the National Natural Science Foundation of China (NSFC) (grant nos. 12374108, 12241401, 12534003, 12204221, and 52271160), the Guangdong Provincial Quantum Science Strategic Initiative (grant no. GDZX2401002), the GJYC program of Guangzhou (grant no. 2024D01J0087), the Beijing Natural Science Foundation (grant no. F252072), and the Beijing Science and Technology Plan Project (grant no. Z241100004224004).






\bibliography{Maintext}{}
\end{document}